# A General Framework for Complex Network Applications

Liu Xiao Fan（刘肖凡）[1], Chi Kong Tse（謝智剛）[2]

(1. School of Computer Science and Engineering, Southeast University, Nanjing, China;
2. Department of Electronic and Information Engineering, The Hong Kong Polytechnic University, Kowloon, Hong Kong)

**Abstract:** Complex network theory has been applied to solving practical problems from different domains. In this paper, we present a general framework for complex network applications. The keys of a successful application are a thorough understanding of the real system and a correct mapping of complex network theory to practical problems in the system. Despite of certain limitations discussed in this paper, complex network theory provides a foundation on which to develop powerful tools in analyzing and optimizing large interconnected systems.

## 1 Introduction

In the past fifteen years, the underlying network structure of complex systems has attracted extensive study from physics and computer science communities. The structural properties of complex networks in engineering infrastructure, social communities, biological systems, and financial systems are closely examined. Important universal properties such as scale-free structure, small-world phenomena, community structure, and dynamical processes are found in complex networks from multiple domains [1]. Efforts have also been made to apply complex network theory to not only describing the topological and dynamical properties of real-world systems, but also to solving practical problem and even re-designing the system for better performance. In this paper, we present a general framework for applying complex network theory in solving real-world problems. First, we review the network construction process of finding the abstract representation of real-world systems. Then, we review the existing analysis of network properties from different scopes. Finally, we discuss the feasibility of using complex network theory to solve real-world problem, including its capability and its limitation.

## 2 Construction of Complex Networks

The fundamental pre-requisition of a successful application of complex network theory is finding the underlying network structure of complex systems. A network is a set of nodes connected by a set of edges. Most complex systems consist of a collection of components which interact with each other. For instance, the Internet is a collection of computational devices connected by wires or wireless signals. Here, the devices are the nodes in the network and the physical connections are the edges in the network. Computers and devices communicate with each other by exchanging data packages. However, the representation of nodes and edges may be more flexible for many complex systems. For example, in the biological system, each species can be viewed as nodes in the network, while the predator-prey relationship and mutual-dependence relationship shape the edges between each species in a food web. In a microscopic perspective, each living organic intake food and generates energy through a chemical process called metabolism. In the metabolism process, chemical substances react with each other and transform into new chemical substances. In the metabolic networks, the nodes are the chemical substances and the edges are the possible transformation from one substance to another. Moreover, different complex systems can overlap and interfere with each other in real-life, forming a network of networks. For example, a social network is a network of people connected by family ties, collaboration and friendships. In modern life, people keep up with friends and maintain their social relationships by using the Internet – a network of computer and smart phones. Furthermore, the complex network of electrical transmission supplies the power that keeps the Internet running. Each of the above mentioned networks are closely coupled with each other. Finding the underlying network structure poses a great challenge yet lays the groundwork of applying network theory to solving practical problems. Generally, the nodes in the underlying network of the complex systems are the physical components, and the relationship of components can be defined in six different ways, as summarized in Table 1 [1].

TABLE I. SIX WAYS OF CONSTRUCTING COMPLEX NETWORKS

| Type of Edges | Typical Networks |
| --- | --- |
| Communication | Email, phone, social, mail |
| Coexistence | Domains, collaborations, books, music, movies |
| Reference | Wikipedia, web, articles, forms, emails, software |
| Confluence | Cities, highways, undergrounds, circuits, power-grid |
| Correlation | Climate, financial market, neuroscience |
| Adjacency (temporal and spatial) | Earthquake, landscape, linguistic |

# 3 Analysis of Network Properties

The properties of complex networks can be examined from different scopes. Here, we categorize the existing analysis of network properties into three scopes, i.e., the macroscopic view, the microscopic view and the dynamical view.

3.1 The Macroscopic View

In the macroscopic view, the statistical properties of the complex networks, such as the degree distribution, the community structure and the structural robustness, are of particular interest.

The degree distributions of complex networks of real systems have shown a very interesting universal characteristic, i.e., they all follow the Zipf law, also known as the power law. Let $k$ be the degree of nodes, the probability $p(k)$ of finding a node with degree $k$ in the network follows

$$p(k) \sim k^{-\gamma}, \qquad (1)$$

where $\gamma$ is known as the power-law coefficient. The power-law coefficients of most complex systems fall between 2 and 3. For example, for the Internet, $\gamma = 2.4$; for the scientific collaboration network, $\gamma = 2.5$ and for the protein-protein interaction network, $\gamma = 2.4$ [2]. The power-law degree distribution reveals the winner-takes-all nature of the complex system. That is, most of the edges in the networks are connected to only a few number of nodes.

Many complex networks are actually loosely connected by several densely connected sub-networks. The sub-networks are called community structure. The detection of community structure in complex networks generally takes two different approaches. The first approach is the "top-down" approach, where algorithms search for the densely connected sub-networks in the network, be them cliques or sets of nodes with maximum modularity. The second approach is the "bottom-up" approach, where specific edges, known as the "weak-ties" are removed from the network, while the remaining disconnected sub-networks are the communities in the network. The weak-ties may refer to the minimum-cut of the network, or edges with largest betweenness centrality [3].

In sociology, the term "assortative", also known as "homophily", refers to the tendency of individuals with similar characteristics, e.g., age, nationality, religion, etc., know or interact with each other. In complex network theory, assortative mixing specifically refers to the bias of preference that nodes with similar degrees are connected together. The opposite term of assortative mixing is disassortative mixing, which refers to the bias of preference that nodes with dissimilar degrees are connected together. Assortativity is commonly observed in social networks. While disassortative mixing exists in biological and technology networks such as the Internet and food webs [4].

3.2 The Microscopic View

The analysis of complex network from the microscopic view focuses on single nodes or the combination of a few number of nodes.

In social networks, there is a likelihood that two friends of a person are also friends themselves. In complex network theory, the clustering coefficient $C$ is a measure of the likelihood of closed triplets, i.e., three nodes that are fully connected. It is defined as:

$$C = \frac{\text{number of closed triplets}}{\text{number of connected triples of vertices}} \quad (2)$$

Clustering coefficient represents the redundancy of edges that keep the network connected. Social networks show large clustering coefficients, for people tend to form a closed society, e.g., family, school, working environment, etc. While in technological networks and infrastructures, the clustering coefficients are small, because the redundant links between nodes increase the cost of the systems [1].

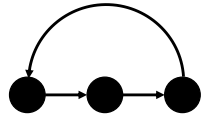

Fig. 1. A motif with three nodes that reproduces a feedback loop.

Network motifs are defined as recurrent and statistically significant small-sized sub-graphs. The network motifs are usually related to the functional properties of the network. For example, a simple motif shown in Fig. 1 reflects a feedback loop, which is a common design in electronic circuits and control systems. Despite of the functional importance of network motifs, their detection is computational challenging. Define $n_M$ as the number of appearance of motif $M$ in the network, $\langle n_M^{\text{rand}} \rangle$ and $\sigma_{n_M}^{\text{rand}}$ are the number of appearance and standard deviation of motif $M$ in multiple randomized networks, then the statistical significance $Z_M$ of motif $M$ is defined as:

$$Z_M = \frac{n_M - \langle n_M^{\text{rand}} \rangle}{\sigma_{n_M}^{\text{rand}}}. \qquad (3)$$

Due to the computational complexity of the algorithms for calculating the statistical significance of a motif, the sizes of motif reported in existing literature are usually limited by 10 [5].

The study of structural properties of individual nodes reveals the importance of a component in the system. The measurements usually take consideration of the ego-network structure of the node. The most straight-forward measure of importance of a node is the number of edges connected to it, i.e., its degree centrality. However, the degree is not a sophisticated measure of node importance in many cases. Other measures based on the structural properties of the ego-network of each node are proposed. For example, the number of indirect neighbors of a node can also be used to extend degree centrality [6]. The distance between a pair of nodes is the length of the shortest path between two nodes in the network. Betweenness centrality of an importance measure of the node. It is calculated based on the number of shortest paths of all pairs of nodes that include this node. The importance of a node sometimes depends on the importance of its neighbors. Based on this idea, the PageRank algorithm is proposed to rank the importance of webpages. The PageRank algorithm, among many other algorithms, are considered related to the eigenvectors and eigenvalues of the adjacency matrix of the complex networks [7].

3.3 The Dynamic View

Most complex systems are not static but rather dynamic. On one hand, the topology of complex networks changes over time. On the other hand, dynamical processes are also taking places on the networks.

A traditional area of study on topological dynamics of complex networks is the robustness of the network. By gradually removing random edges from the network, a strongly connected network may transform into several unconnected sub-networks. The critical proportion of edges removed in order to disconnect the sub-networks reflects the topological robustness of the network. Study has found that real networks with scale-free structure display great robustness against random edge removal. However, the real networks are more vulnerable to removal of important, rather than random, nodes and edges. The strongly connected network can be quickly disconnected into several sub-networks [8].

Complex networks play a crucial role in carrying contents and facilitating communications. For example, information spreads on the Internet through social networking services, disease and behavior spreads in social community forming epidemics, etc. Understanding the mechanism of content spreading is the foundation of predicting epidemic spreading and identifying super spreaders. The traditional model of epidemic spreading is SIR (susceptible-infected-recovered) model. This model assumes that a population can transform with a certain probability from the susceptible state to the infected state and from the infected state to the recovered state. The SIR model is a simplified model of epidemic scenario. Many other models, including SIRS (susceptible-infected-recovered-susceptible) model, SEIR (susceptible-exposed-infected-recovered) model, etc. Content spreading on complex networks takes similar form to the epidemic spreading in social community.

One of the ongoing discussion in theoretical complex network study is the adaptive co-evolution of network topology and dynamical processes. On one hand, the underlying network structure strongly affects the dynamical process such as communication and epidemic spreading. On the other hand, the dynamical process may also alter the topological structure of the complex networks. Up to now, the problem has been tackled from several angles, such as game theory on network models, self-organization networks and opinion formation in social networks. However, the adaptive co-evolution will pose a continuous challenge to network scientists [9].

# 4 Solving Real-world Problems

In the previous sections, we have reviewed the construction methods of complex networks and their fundamental properties. In this section, we propose a general methodology of applying the theory to solving real-world problems. The key of a successful application is the correct mapping of network properties to practical problems. Finding such a mapping requires an in-depth understanding of the real system as well as a systematical knowledge of network science. Here we outline some typical systematical problems that are particularly suitable being solved by complex network theory.

4.1 Re-discovering System Structure

An epidemic model is a simplification of disease or behavior spreading. The epidemic threshold (or reproduction number) refers to a certain probability that an epidemic occurs only if the infect probability of the disease or behavior is larger than the epidemic threshold. In the SIR model, assume $\beta$ the spreading rate, i.e., the transform probability from susceptible state to infected state and $\mu$ the removal rate, i.e., the transform probability from infected state to recovered state, in order to ensure an epidemic outbreak, the following condition has to be met:

$$\frac{\beta}{\mu} \geq \frac{\langle k \rangle}{\langle k^2 \rangle - \langle k \rangle}, \qquad (4)$$

where $\langle k \rangle$ is the average degree of nodes in the underlying transmission network. In traditional epidemic research, social communities are considered as fully connected networks or random networks. In these networks, $\langle k \rangle / (\langle k^2 \rangle - \langle k \rangle) \gg 0$. Therefore the epidemic threshold exists. However, recent study has shown that human contact networks are neither fully connected networks nor random networks, but rather scale-free networks (or at least networks with long-tail degree distribution). In scale-free networks, some hub nodes can have very large degree, hence $\langle k^2 \rangle \to \infty$ and $\langle k \rangle / (\langle k^2 \rangle - \langle k \rangle) \to 0$. In this case, the epidemic threshold does not exist, and that the disease may have a break out even if the infectious probability is low [10]. The discovery of scale-free property of the social community has fundamentally changed the understanding of immunization strategy. New immunization policies have been proposed in order to accommodate to the change [11].

4.2 Partitioning and Categorizing System Components

Public companies are traded in the stock markets. The companies are usually categorized into sectors by their nature of business, e.g., real estate sector, financial sector, technology sector, etc. Spreading investments across different sectors are believed to decrease the systematic risk of the portfolio. However, the existing sectoring criteria are sometimes insufficient since modern companies tend to diverse their business into different sectors. A robust sectoring method is required in modern investment activities. One of the solutions is the community detection algorithms in complex network theory. First, a stock market network must be constructed. The nodes in the network are traded companies. Every pair of the nodes are connected by an edge. The weight of edges are given by the correlations of the time series of stock returns. The resulting stock market networks show clear community structure. The compartments of stock markets sectored by community detection algorithms are basically consistent with traditional sectoring methods but also provide additional insights into the difference among companies within a same traditional sector. The sectoring method also shows stronger flexibility on the resolution of market sectoring [12].

4.3 Importance Ranking of System Components

We can find the need of ranking components in the system for practical usage in many scenario. For example, a user usually only reads the first two or three results returned from a search engine, an advertiser can only afford advertisement in one or two influential spreaders in social media websites, etc. By modeling the complex systems to networks, the importance ranking of individual components can be revealed by its ego-network structure. PageRank algorithm relates the importance of webpages to the eigenvector of the underlying network of the World Wide Web. Epidemic models are used to find influential spreaders in social networks [13]. In the fight against terrorism, critical information carriers are identified by calculating the betweenness centrality of each node in the terrorists' social networks [14].

4.4 Recovering Missing Information

Complex networks are built from observation data. However, the data collection process can be compromised by imperfect technology or human error, resulting in incomplete data or faulty data. Therefore, recovering missing knowledge from existing information is of urge need in many practical problems. For example, recommender systems use data on past user preferences to predict possible future likes and interests. By building a bipartite network of user and objects, the structural similarities between different users or different objects can be calculated. Accurate and diverse recommendation can be made by correctly associating users with potential objects purchased by similar users. Similar methodology can be applied in prediction of protein functions and inference of latent terrorist's relationships [15].

4.5 Designing Bionic Systems

Although scale-free networks are robust to random failure, an error happened on its important nodes can cause cascading failure that could potentially sabotage the whole system. For example, the failure of a highly connected line on the power grid will redirect electrical power to other lines that may not have the capacity to handle the increased power, creating a regional black-out. On the other hand, social network is an example of continuously evolving system that exhibits strong robustness even to attacks on the most important nodes. Similar to social networks, the swarms of fish and flocks of bird also possess the ability of self-organizing that can adjust stability or maintain synchrony of the system in real-time. Electrical engineers have already started to transform the ideal synchronization models and self-organizing models learnt from complex network theory to the engineering models of power grids. However, the gap between physics and practical engineering is still very large and yet to be filled [16].

# 5 Limitations of Applied Network Theory

Despite of the fruitful applications of complex network theory in solving many practical problems, there are certain limitations of the tools. Particularly, the limitations are on the oversimplification in modeling complex systems with networks.

For example, the power grid is the largest and most complicated infrastructure. Traditional complex network analysis of power grid treats generator, user and voltage transformer as network nodes, while transmission lines as edges. The electrical power are modeled as network flow carried by the underlying complex network. However, electrical engineers have criticized this model as oversimplified. Actually, the practical mathematical models used by the engineers are highly nonlinear and difficult to analyze. Although synchronization models on complex networks, such as Kuramoto model, parallel with the power grid in many aspects, there is still much work to do in order to apply complex network theory in optimizing of electrical transmission systems.

Another example is discovering the social circles in ego networks. In social networks, people are connected by multiple types of social relationships, e.g., colleagues, friendship, family ties, etc. Discovering the nature of ties between users is an important challenge for online social services providers. It has been found that community detection algorithms that use merely the topological properties are not sufficient in inferring the correct social circles of users. Other features, such as age, geographic location, education background, etc., should be used together with topological information of the social network to achieve accurate results [17].

# 6 Conclusion

In this paper, we have presented a general framework of applying complex network theory to solving practical problems. The fundamental step of the application is the correct modeling of real systems into networks. By analyzing the network structural properties and mapping the structural properties to the functionality of real systems, complex network theory can be applied to revealing of importance of system components, identifying compartments in the system, predicting system behavior and even redesigning the system to achieve better performance and robustness. However, a successful application has to meet challenges in many aspects. First, correctly modeling a real system to networks and finding the mapping of network property to practical problem require an in-depth understanding of the real system as well as a comprehensive knowledge of complex network theory. Second, oversimplified network models may not fully characterize the evolving mechanism of the real system. Tools from other academic fields, such as nonlinear theory, machine learning algorithms, etc. should all be utilized along with complex network theory to explore solutions to practical problems.

# Acknowledgement

Xiao Fan Liu is supported by the Natural Science Foundation of China (Nos. 61304167, and 61374170) and the Natural Science Foundation of Jiangsu Province (No. BK20131301).